# Design and electron optics performance of a MEMS electrostatic electron monochromator

***M.J. Adriaans[1], J. P. Hoogenboom, A. Mohammadi-Gheidari[2]***

**Department of Imaging Physics, Delft University of Technology, Delft, The Netherlands**

**E-mail addresses:** [1]m.adriaans@tudelft.nl, [2]a.m.gheidari@tudelft.nl

**Keywords:** Electrostatic monochromator, Fringe-field deflector, MEMS fabrication, Low-voltage SEM (LVSEM), Energy Filtering, Electron optics

Abstract

Monochromators are routinely used in Transmission Electron Microscopy and Electron Energy Loss Spectroscopy, to improve both spatial and energy resolution. State-of-the-art monochromators, however, are complex instruments that typically require additional electron optical correctors, limiting their implementation to the high-end, most expensive microscopes. Miniaturized monochromation relying on purely electrostatic fringe fields has recently been proposed as a means to realize a simpler and thereby more cost-effective and easier to operate high-resolution monochromator. Here, we present a design for such a compact, fully electrostatic fringe-field based monochromator. Our design consists of a superposition of an Einzel lens with a series of electrostatic deflectors and is entirely based on dimensions that can be realized with MEMS fabrication technology. Thanks to mild mechanical and electrostatic potential tolerances, the MEMS-fabricated electrodes can be passively aligned and only need seven regular power supplies. We present an analysis of spectral broadening due to diffraction in our monochromator design as well as particle simulations including stochastic Coulomb interactions. This analysis shows that our design can achieve a resolution of 19 meV while maintaining 128 pA of beam current and thus potentially achieve energy filtering comparable to state-of-the-art monochromators. Our MEMS monochromator could therefore bring the application of energy filtering into the domain of SEM and specifically allow higher resolution imaging in chromatic-aberration dominated low-voltage SEM.

# 1 Introduction

In Scanning Electron Microscopy (SEM), low (< 5keV) electron beam landing energies are used in applications ranging from biological imaging to semiconductor inspection and metrology [1], [2], [3], [4], [5], [6]. At lower beam energies, electron beam interactions are confined to a smaller volume compared to higher energies, reducing the volume in which signal is generated and thereby increasing resolution. For topographic imaging or other applications where detection of secondary electrons (SEs) is beneficial, an additional advantage is the higher SE yield at lower energies. Moreover, a lower back-scattered electron (BSE) yield and BSE energy may reduce the amount of BSE-generated SEs in the total SE signal and thus lead to higher signal to background. For insulating or poorly conducting samples, low-voltage SEM (LVSEM) may have the further benefit that the total electron yield is close to unity, mitigating charging artefacts.

A common challenge for LVSEM applications is the fact that lens aberrations dominate the achievable resolution. Especially the contribution of chromatic aberrations, scaling with the relative energy spread of the beam and the beam opening angle, limits the resolution obtainable in LVSEM. Novel electron source concepts, with potentially reduced energy spread, have been proposed[7] but so far not found implementation in commercial SEMs. An alternative approach is to use an existing high-brightness source and either monochromate the beam further down the beam path or compensate the induced aberrations using aberration correctors. Electron monochromators and aberration correctors have found widespread application in (Scanning) Transmission Electron Microscopy ((S)TEM) and Electron Energy Loss Spectroscopy (EELS), where they have enabled atomic-scale spatial and few-meV energy resolution [8], [9], [10], [11], [12], [13].

These state-of-the-art monochromators and aberration correctors are complex electron optics instruments that require many individual components with extremely tight electro-mechanical tolerances. The latter can typically not be met by current fabrication techniques requiring additional corrector elements adding costs and complexity. As a result, monochromators and aberration correctors are typically only used in the most expensive high-end, high-resolution applications of (S)TEM. Miniaturized correctors and simplified monochromator concepts have been proposed to scale down cost and increasing useability [14], [15], [16].

Recently, we have proposed a new concept for electron beam monochromation that could specifically be implemented with a miniaturized device[17]. In this concept, purely electrostatic fringe fields are used to create the necessary dispersion. In this manuscript, we present a design for a miniaturized, MEMS-based monochromator relying on electrostatic fringe field dispersion. A thorough electron optics analysis and optimization shows that our design can achieve meV resolution, without relying on extremely tight tolerances or additional correctors. The remainder of this manuscript is organized as follows: First, the general requirements for this system for operation in LVSEM are described, leading to a minimal mechanical layout of the electrodes in the design. Next, for an optimum performance, the voltages applied to the electrodes are automatically optimized. The performance of the monochromator is then evaluated through simulation of the filtered energy spectrum. Finally, the main electric and mechanical tolerances of the monochromator are discussed.

# 2 Basic design considerations

## 2.1 Design requirements

For this new MEMS monochromator to be practically useful in a SEM, there is a number of general requirements regarding performance and operability. These are listed and defined in this section.
First the achievable filtered current should be high enough for user friendly focusing and smooth imaging in a SEM. This, in turn, requires optimal tuning of the electron beam passing through a dispersive electrostatic deflector, to optimize between current and energy resolution. This optimal relationship was analyzed in our previous manuscript and is described by energy resolution limit $d\phi_{\alpha,\mathrm{Co,B}}$ [17], which is given by

$$d\phi_{\alpha,\mathrm{Co,B}} = 2I_0\left[\left(\frac{Km}{\epsilon_0 e^2}\right)^2\left(\frac{12}{B_r\pi^2}\right)\right]^{\frac{1}{5}}, \tag{1}$$

where $B_r$ is the reduced beam brightness, $\epsilon_0$ is the vacuum permittivity, $e$ is the electron charge, $m$ is the electron mass, $I_0$ is the unfiltered beam current and $K = 0.642$ is a numerical constant that describes the magnitude of Boersch energy broadening in a Pencil beam regime [18]. The monochromator resolution should thus approximately achieve $d\phi_{\alpha,\mathrm{Co,B}}$. Second, the electron beam should enter and exit the monochromator along a straight optical axis for straightforward integration into the SEM column. Third, the mechanical tolerances of the electrodes should allow for passive alignment, accurate to at least 10 micrometers [19], [20]. Fourth, for operability and ease of construction, the design should avoid magnetostatic components and use as few electrodes and voltage supplies as possible. Fifth, the monochromator resolution should be continuously tunable between unfiltered, switched-off, maximum current use to the optimally filtered, maximum energy resolution, use. This way, the operator can easily switch from relatively high current inspection to using the lower current monochromated beam while the operating settings of the microscope around the monochromator remain independent of working status of the monochromator. Finally, unlike a homogeneous deflector based monochromator, the voltage stability requirement should not be smaller than $d\phi_{\alpha,\mathrm{Co,B}}$ such that this requirement does not exceed that of other electrodes typically present in a monochromated LVSEM after the monochromator.

## 2.2 General layout and dimensions

A mechanical design that fulfills these general requirements, is introduced next. The layout of this design is schematically depicted in *Figure 1*. Central in our design is a main dispersive deflector, denoted by D1. D1 provides the dispersion that ultimately allows filtering of the electron energies with a final selection aperture. The coordinate system is defined such that electrons enter the monochromator along the $z$-axis, and D1 bends the beam in the negative $x$ direction. Thus, higher energy electrons are dispersed in the positive $x$-direction compared to lower energy electrons. To increase energy dispersion, the beam is decelerated before reaching D1 and accelerated immediately after. This combined deceleration and acceleration effectively forms an Einzel lens into which D1 is immersed (see also *Figure 1*). This Einzel lens then also focuses the beam towards the selection aperture. A crucial feature of this configuration is that electron-electron interactions are maximally confined to the region where the beam is deflected by D1, which is needed to approach the limit $d\phi_{\alpha,\mathrm{Co,B}}$. To form the fields of the Einzel lens, cylindrically symmetric electrodes, referred to as L1 and L2, are added immediately before and after D1 and, together with D1, are biased relative to another pair of grounded circular electrodes denoted L0 and L3.

For the beam to enter and to leave the monochromator along the original $z$-axis, an additional pair of deflectors is required for beam alignment. These deflectors, denoted D0 and D2, can either be immersed in the decelerating/accelerating fields between L0-L1 and L2-L3 respectively, or be placed outside these fields before L0 and after L3. By immersing D0 and D2 in the decelerating and accelerating fields the beam is kept closer to the optical axis. In contrast, placing the deflectors outside of the decelerating/accelerating fields would require additional grounded shielding electrodes to prevent the fields from extending towards the selection slit. For example, a deflection field of 100 V/mm across a 100nm selection slit corresponds to a voltage variation of approximately 10 mV, that is shielded by the shape of the conductive selection slit itself, which is difficult to control. Therefore, deflectors D0 and D2 are immersed in the fields between the grounded and biased electrodes of the Einzel lens.

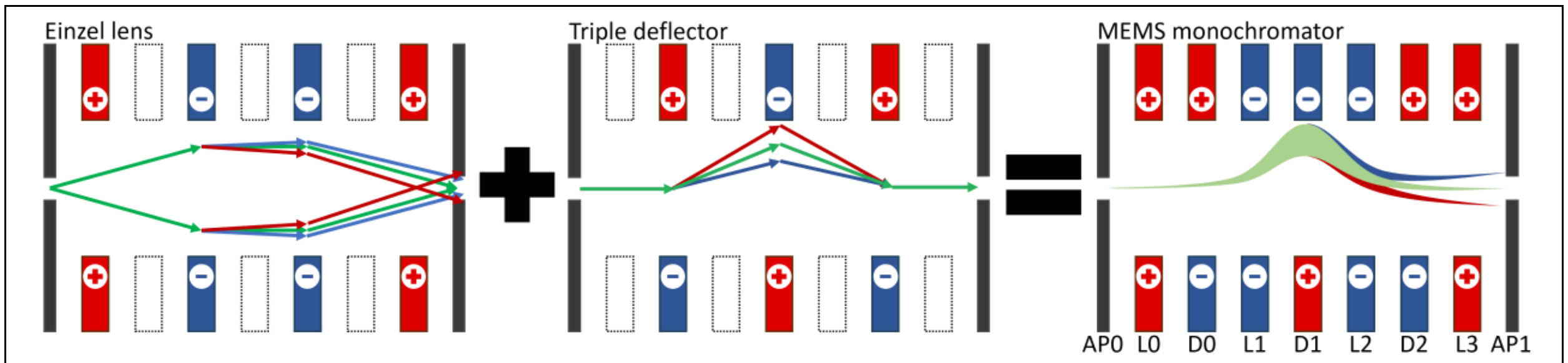

*Figure 1: Schematic illustration of the electrodes in the MEMS Monochromator (right). The monochromator consists of a superimposed retarding Einzel lens (left) and a triple deflector series (middle). Positively charged electrodes are indicated in red, negatively charged electrodes in blue. AP0, AP1: input and output plane apertures, AP0 serves to filter the tails from the initial input electron distribution; L0, L3: grounded lens electrodes; L1, L2: biased lens electrodes; D0: initial deflector steering the beam into D1: the main dispersive deflector; D2: final deflector steering the beam back to the central axis.*

Two apertures complement the design. The input aperture, AP0, is used to filter the tails of the entrance beam distribution at initial focus, and the second aperture, AP1, is used to select a narrow range of energies from the dispersed spectrum. Assuming a round energy selection aperture, its diameter, typically < 1 µm, will be smaller than any feasible passive mechanical alignment accuracy. Therefore, an additional deflection field is introduced in the plane of D1 which can steer the beam in the y-direction. This is achieved by adding 2 more small electrodes to the deflector, together forming a quadrupole as depicted in *Figure 2*b. The quadrupole can be excited to correct for astigmatism, without requiring additional electrodes or voltage channels and without loss of prior mentioned functionality.

Central to the monochromator concept is the use of wide, overlapping fringe fields. Moreover, for an adjustable energy resolution, the opening of D1 should be wide enough for the beam to pass through at a large deflection angle, but also without being deflected (continuing through the monochromator without dispersion). A symmetric, wide opening would, however, unnecessarily increase the risk of electrical discharges at large deflection angles, since the attracting electrode of D1 would have to be at a high voltage relative to the surrounding electrodes. This risk can be reduced by shifting the deflectors off-center, which also causing a beam maximally deflected in the dispersion direction to pass closer to the center of the y-deflector. Thus, this shift should minimize the additional deflection aberrations when deflecting the beam in the $y$-direction. Therefore, the deflectors are intentionally positioned off-center with respect to the common $z$-axis of the system.

Since the design is based on short fringe fields, MEMS fabrication techniques are used to accurately shape the electrodes, which means that the dominant tolerances are primarily determined by assembly and stacking misalignments. The choice for MEMS fabrication also means that the thicknesses are limited to typical wafer thicknesses of 300 µm and 525 µm. Since the deflection coma scales inversely with the deflector length [17], the deflectors are designed to be thicker (525 µm) than the other electrodes (300 µm). While this does not directly influence the theoretically achievable energy resolution, it shifts the optimal operating conditions toward higher beam energies and larger beam spot radii which are typically easier to realize experimentally. Finally, the equal radii of L0 and L3 are chosen to be smaller than the other geometrical dimensions in the design. This creates a rotationally symmetric lens effect despite the nearby fields generated by D0 and D2. This also shields fields from the other electrodes unfavorably interacting with the selection slits and its supporting structures. The design fulfilling these requirements is shown with accurate proportions in *Figure 2* and the corresponding dimensions are listed in Appendix 6.1.

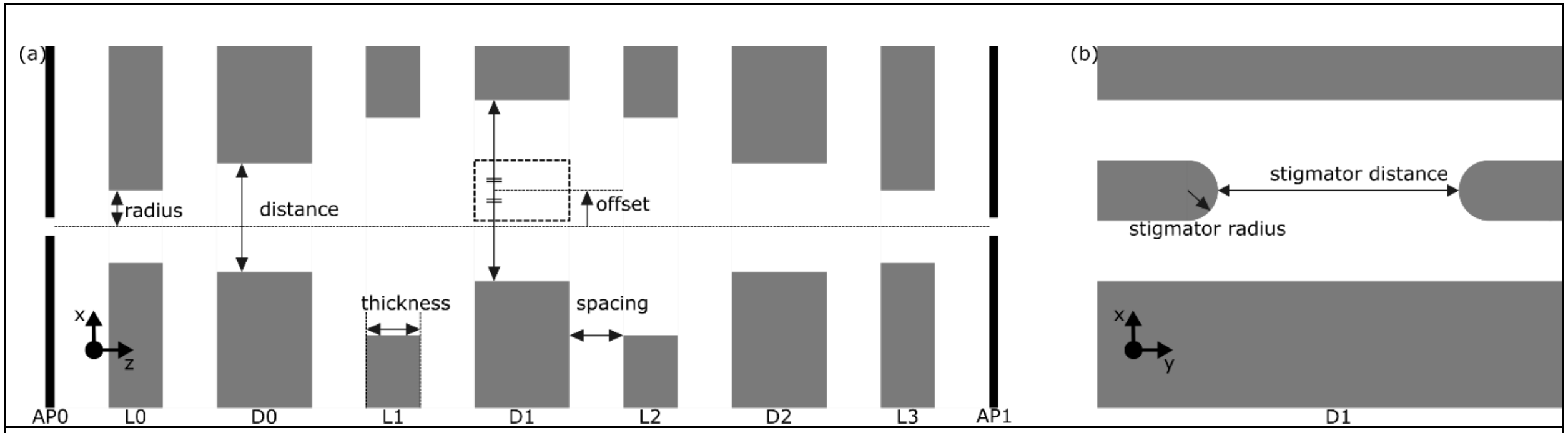

*Figure 2: Optically critical dimensions of the electrodes used in the design. (a) Cros-section in xz through the center $y = 0$. The sketch corresponds to relative proportions, except for the radii of AP0 and AP1. L0, L1, L2 and L3 indicate the positions and sizes of the round electrodes and D0, D1 and D2 indicate the positions and sizes of the deflector plates. The dashed rectangle in the D1 plane indicates the location of the stigmator tips. (b) Cross-section in the xy-plane through the dashed rectangle in (a) showing the stigmator.*

These mechanical dimensions are not fully optimized to reach the best possible performance, but are taken as a constant in determining the remaining free design parameters. Similarly, the size of AP0 and AP1, influence the performance of the monochromator and should therefore be decided. The required current is calculated first and is used to determine the other free parameters of the beam. For a reasonably good signal/noise imaging, assuming at least 100 electrons in $10^6$ sampled pixels at 10 frames per second, a filtered current of approximately $I_1 = 160$ pA is deemed reasonable. In an ideal situation, the filtered current $I_1$ depends on the unfiltered current $I_0$ as

$$I_1 = K_d \frac{\mathrm{FW}_{50,1} I_0}{\mathrm{FW}_{50,0}}, \tag{2}$$

where $\mathrm{FW}_{50,0}$ and $\mathrm{FW}_{50,1}$ indicate the full width that contains 50% of the total energy distribution before and after the monochromator respectively. This expression assumes a uniform initial spatial intensity distribution with a uniform spectrum being dispersed passing through an aperture of the same radius as the initial source. The numerical factor $K_d = 1.19$ is obtained by considering the overlapping area A, adapted from [21], between 2 disks

$$\mathrm{A(x)} = 2\mathrm{R}^2 \cos^{-1}\left(\frac{|x|}{2R}\right) - \frac{|x|}{2}\sqrt{4R^2 - x^2}\,, -2R < x < 2R, \tag{3}$$

where $x$ is the distance between the centres of the disks and $R$ is the common radius. Integration and normalization to find the $\mathrm{FW}_{50}$ for this distribution A yields the factor $K_d$. Assuming an achievable resolution of $\mathrm{FW}_{50,1} = d\phi_{\alpha,\mathrm{Co,B}}(I_0)$, with $\mathrm{FW}_{50,0} = 1$ eV and $B_r = 10^8$, an input current of $I_0$=11 nA is required. The length of the deflection field $l_0 \approx 2$ mm (twice the deflector distance of D1), leads to an optimal potential at D1, $\phi_{\mathrm{D1}}$, using [17]:

$$\phi_{\mathrm{D1}} = I_0 l \left(\frac{B_r \pi^2}{12}\right)^{\frac{1}{5}} \left(\frac{Km}{\epsilon_0 e^2}\right)^{\frac{3}{5}} \approx 112 \text{ V}. \tag{4}$$

The corresponding optimal internal beam angle at the deflector $\alpha_{\mathrm{D1}}$ is

$$\alpha_{\mathrm{D1}} = \left(\frac{3I_0}{2B_r \pi^2 \phi_{\mathrm{D1}} l_0^2}\right)^{\frac{1}{3}} \theta, \tag{5}$$

where deflection angle $\theta \approx 1$ rad. The source and image radius are then approximated by $r_0 \approx \frac{\alpha_{\mathrm{D1}} l_1}{2} = 76$ nm, where $l_1$ is the total length of the monochromator. However, $r_0$ is rounded to 100 nm, biasing towards a large geometric image size and lower opening angle and hence low brightness loss and improved tolerances, at the cost of energy resolution. Moreover, a smaller free-standing aperture will become increasingly difficult to fabricate reliably. The initial acceleration voltage $\phi$ has to be a factor higher such that deceleration to $\phi_{\mathrm{D1}}$ focusses the beam onto AP1. This value is fixed to 500 V and the other voltages are optimized. The input beam opening angle $\alpha$ is constrained by the other parameters as $\alpha = \sqrt{\frac{I_0}{B_r \pi^2 r_0^2 E}} \approx 1.5$ mrad. For these parameters, $d\phi_{\alpha,\mathrm{Co,B}} \approx 0.01$ V, which will be used for optimizing the voltages. These analytically calculated parameters should thus yield an optimal balance for current and energy resolution, which is evaluated after optimizing the voltages applied to the electrodes.

# 3 Geometric electron optics considerations

The design presented above contains 7 unique voltages that should be set in order for the monochromator to work. For optimally aligned electrodes, $y$-deflection is not needed, requiring optimization of 6 remaining voltages. For this optimization we consider several competing effects, including geometric aberrations, Boersch energy broadening, dispersion strength, and electrical discharge risk between electrodes. In this section, the optimization procedure balancing these effects is described.

## 3.1 General computational approach

The electron optics performance of our design cannot be determined analytically due to the complexity of the fields and electron trajectories. Instead, to optimize the electron optics of the monochromator, we used GPT+BEM [22] for meshing the electrostatic field computation and ray tracing. The mesh relies on several parameters to control the accuracy of different features of the mesh near the path of the beam, leading to a mesh with 0.3 million triangles. With this meshing, the square root of the area of the triangles along the edges of the electrodes near the beam is in the order of $10^{-5}$ µm, which is the required alignment tolerance. Mesh triangles further away are larger, allowing for fast field calculations. The $y$-displacement of the beam after optimization, indicated in Table A3 as a measure of accuracy of the fields along the beam trajectory, is better than the size of these triangles.

The fields are computed prior to ray tracing as a set of surface charge densities that lead to a unit potential per electrode, which are linearly independent solutions. These charge densities are then scaled to the applied voltage such that the electric fields can be flexibly tuned in our iterative optimization. BEM computes these surface charge densities as a matrix inversion iteratively, which is truncated when the difference between the target potential and the potential generated by the charges reaches a relative root mean square precision set to $10^{-5}$, which is in the order of the power supply drift requirement.

GPT is then used to trace 40 sample trajectories which uniformly fill the earlier defined initial beam shape. The output is used for fast characterization of the performance based on fitted geometric aberrations including dispersion, an analytical formulation for Boersch energy broadening, and electrical discharge. The voltages are initially set such that the path of the rays focus while visually resembling the path displayed in *Figure 1*. Then, we use a Gauss-Newton method to optimize a cost vector $\boldsymbol{b}$ that is constructed based on aforementioned effects. This procedure relies on the pseudoinverse of the Jacobian that consists of sensitivities $\boldsymbol{S}_i$ that are computed as

$$\boldsymbol{S}_i = \frac{\partial \boldsymbol{b}}{\partial \phi_i}, \quad (6)$$

where $\partial\phi_i$ represents a small variation in normalized voltage with index $i$. The elements in cost vector $\boldsymbol{b}$ are normalized such that after optimization, the largest contributors and $|\boldsymbol{b}|$ should approach 1. The cost vector contains normalized components $b_{jklmcr}$ which represent aberrations, $b_B$ for Boersch broadening and $b_d$ for discharge risk which will be discussed next section.

## 3.2 Aberration and dispersion contributions

The aberration contributions are defined using aberration function, $\chi$, evaluated far away from the entrance aperture, similar to [13], [23] as

$$\chi = \sum_{j\le k, l\le m, c} \chi_{jklmc} = \sum_{j\le k, l\le m, c} \mathrm{Re}\{-C_{jklmc}\omega_0^j \overline{\omega_0}^k w_0^l \overline{w_0}^m \kappa^c\} = \sum_{j\le k, l\le m, c} \mathrm{Re}\{-C_{jklmc}P_{jklmc}\}, \quad (7)$$

where indices $j, k, l, m$ and $c$ indicate the polynomial order for aberrations with coefficients $C_{jklmc}$, $\omega_0 = x_0' + iy_0{}'$, is the initial complex angle, $w_0 = x_0 + iy_0$ is the initial complex position, $\overline{\omega_0}$ and $\overline{w_0}$ are the complex conjugates of $\omega_0$ and $w_0$ respectively and $\kappa = \frac{E-E_0}{E_0}$ is the relative chromatic offset.

The ray's position displacement at the exit plane, $w_1$ is obtained from the complex derivative of $\chi$ with respect to $\overline{\omega_0}$

$$w_1 = -2\frac{\partial \chi}{\partial \overline{\omega_0}} = \sum_{j\leq k, l\leq m, c} -2\frac{\partial \chi_{jklmc}}{\partial \overline{\omega_0}} = \sum_{j\leq k, l\leq m, c} w_{1,jklmc}, \quad (8)$$

which maps rays from the entrance of the monochromator to the exit plane. The individual ray offsets (dropping image plane index 1) are then given by

$$\begin{aligned} w_{jklmcr,n} &= C_{jklmcr}\left(k\omega_{0,n}^{j}\overline{\omega_{0,n}}^{k-1}w_{0,n}^{l}\overline{w_{0,n}}^{m}\kappa_n^c i^r + j\overline{\omega_{0,n}}^{j-1}\omega_{0,n}^{k}w_{0,n}^{l}\overline{w_{0,n}}^{m}\kappa_n^c i^{-r}\right) \\ &= C_{jklmcr}M_{klmcr,n}, \end{aligned} \quad (9)$$

where $M_{klmcr,n}$ is the corresponding polynomial evaluated for ray $n$, and index $r = 0$ and $r = 1$ correspond to the real and imaginary parts of $C_{jklmc}$, respectively. The aberration coefficients are obtained using a least-squares fit by multiplying the pseudoinverse [24] of the polynomial matrix $M$, containing all elements $M_{klmcr,n}$ , with the simulated image-plane ray positions. The included geometric aberrations range up to order $j + k - 1 = 3$.

To estimate the blur contribution of each aberration term, a single representative ray with $w_0 = r_0$, $\omega_0 = \alpha$, and $\kappa = \frac{d\phi_{\alpha,\mathrm{Co,B}}}{\phi_0} \approx 1 \times 10^{-5}$ is used to compute a signed blur $w_{jklmcr}$ (dropping index $n$ for a single ray) for a given aberration. The resulting signed blur values are then normalized to construct signed optimization cost terms. The dispersion contribution is defined as $b_{010010} = r_0/w_{1,010010}$, while all remaining aberration contributions are normalized as $b_{jklmcr} = w_{jklmcr}/r_0$.

## 3.3 Boersch broadening contribution

The contribution due to Boersch broadening is approximated by integrating the electron-electron interactions along the optical axis. In the pencil beam regime, adapted from [18], the beam gains additional an energy spread $\Delta\phi_\mathrm{p}$ given by

$$\Delta\phi_\mathrm{p} = \sum \frac{Km}{\epsilon_0 e^2}\frac{I^2 \Delta s}{\phi_0}, \quad (10)$$

where $\Delta s$ is the average off-axis distance traveled by the beam between two recorded trajectory frames. The normalized contribution to the cost vector is $b_B = \frac{\Delta\phi_\mathrm{p}}{d\phi_{\alpha,\mathrm{Co,B}}}$.

## 3.4 Electrical discharge

To reduce the risk of electrical discharge, an additional term is included. The electrodes are expected to spark at electric field strengths around $10^7$ V/m [25], [26], [27]. In order to incorporate this effect in $\boldsymbol{b}$, an optimization cost is defined as

$$b_d = \sum \left(\frac{\Delta V_{ab}}{d_{ab}E_{\max}}\right)^4, \quad (11)$$

where $\Delta V_{ab}$ is the potential difference between electrodes $a$ and $b$, separated by a distance $d_{ab}$, and $E_{\max} = 8 \times 10^6$ V/m. With this steeply rising function, the contribution is effectively zero when the electric field strengths are not close to $10^7$ V/m, while stronger fields contribute a very high cost (larger than 1).

## 3.5 Qualitative optimization result

Though the optimization is initialized close to a trajectory such as displayed in *Figure 1*, the optimization process may converge to a different solution. Therefore, the optimization result is qualitatively examined here.

The voltages that result from this optimization are in Appendix 6.2, and the aberrations fitted when applying these voltages are in Appendix 6.2. The voltage are all of the same order of magnitude, which helps in later experimental implementation.
The costs combine to a magnitude $|\boldsymbol{b}| \approx 2.0$, indeed in the order of unity required to approach a resolution of $d\phi_{\alpha,\mathrm{Co,B}}$. The resulting beam trajectory is qualitatively shown in *Figure 3*, bending and focusing as expected. As is visible in the figure, D0 initially induces dispersion with a negative sign which is present throughout deflector D1, which is then more than compensated by D1, leading to a dispersed image. However, this is not yet a quantitative result for the energy resolution, which is computed next.

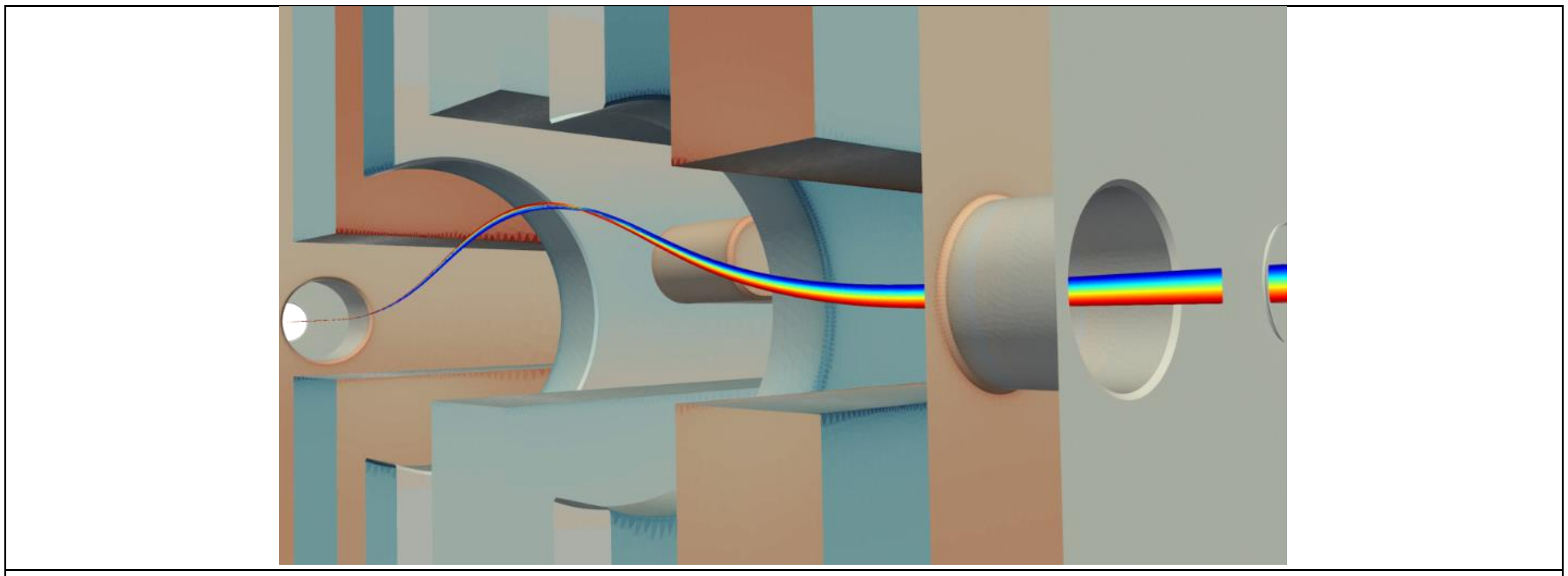

*Figure 3: Simulated beam trajectory after optimization. The beam enters through an aperture at the virtual source plane on the left side of the image and is focused and dispersed in the image plane on the right side of the image. A cross section of the electrodes that focus and bend the path of the electrodes are shown as well. The colored surfaces of the electrodes depict the surface charge densities, and the rainbow-colored beam depicts the trajectories of a particle beam with an exagerrated initial energy spread and opening angle. The negative dispersion caused by D0, which is visible in the figure when the beam passes D1, is more than compensated by D1.*

# 4 Monochromator performance

## 4.1 Particle interactions and energy spectrum

A monochromator disperses electrons spatially according to their energy, thus creating a correlation between transverse position $x$ and energy $E$. Including the effect of chromatic defocus at both tails of the initial energy distribution this produces a dispersed spot shaped as indicated in *Figure 4*. By selecting a narrow spatial slice at the energy-selection plane, the energy spread of the transmitted beam is reduced. The projection of this slice of the particle distribution on the $E$ axis yields a density spectrum, which can be used to evaluate the monochromator performance. Two quantities that describe the density spectrum are the $\mathrm{FW}_{50}$ and the Kimoto limit, $\mathrm{FW}_{\mathrm{K}}$ , corresponding to the full width at 50 % of the maximum intensity and the full width at 0.1% of the maximum intensity, respectively. The transmitted spectrum is computed for the monochromator with the optimized voltages applied to the electrodes by tracing $10^7$ electrons through the monochromator geometry like described above but now including stochastic electron-electron interactions explicitly. A single virtual aperture could be used to collect a portion of the initial current, leaving a very small fraction of the current to form the spectrum. Instead, to acquire sufficient sampled electrons for computing the $\mathrm{FW}_{\mathrm{K}}$, more virtual apertures are used to collect a larger fraction of the current, as is schematically illustrated in *Figure 4*.

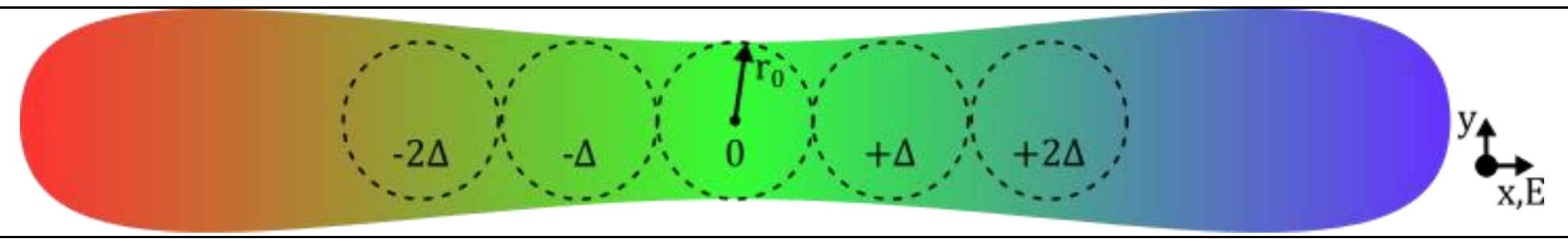

*Figure 4: Schematic illustration of the energy-dispersed beam profile at the energy selection plane including chromatic defocus. Dashed circles indicate the position of virtual apertures with radius $r_0$ used to calculate energy spectrum and fraction of beam current after filtering. The central virtual aperture represents the nominal beam energy. For evaluation of the monochromator performance, we sample with more apertures, which is possible as long as they are close enough to the central beam. The discrete energy offsets relative to the nominal beam energy are indicated in units of Δ per aperture.*

The depicted dispersion pattern has a central region along the $x$-direction in which the chromatic defocus remains sufficiently small such that the spot size for a given energy changes negligibly. Provided that the initial energy distribution is significantly broader than the monochromator resolution and the intensity of the initial spectral density remains approximately unchanged along this central portion of the dispersion pattern, multiple virtual apertures can be placed at different $x$-positions, where the spectral lineshape is not significantly affected. Therefore, the initial energy distribution is assumed to be uniform with a width of $2\mathrm{FW}_{50} = 2$ eV [28]. Each virtual aperture corresponds to a different nominal energy, which differs by discrete amounts per aperture equal to $\Delta = 2r_0/C_{010010}$. This is corrected for by subtracting this offset from each collected electron before binning the electrons to form a spectrum.

The beam dimensions that are analytically calculated using Equations 4 and 5 should approach the theoretically optimal monochromator tuning. While a more accurate generalized optimum could be reached by further iterating over these and other mechanical parameters, it should be noted that a different optimal tuning would apply depending on $B_r$, initial energy spread, and desired output current.

Therefore, to investigate whether these parameters indeed yield optimal current and energy resolution, the opening angle $\alpha$ is varied from 0.5 to 2.5 mrad while maintaining a constant reduced brightness of $B_r = 10^8$ A/(m$^2$ sr V), thereby changing the beam current. The brightness loss is defined as $L \coloneqq 1 - T_d$, where the distribution transmission $T_d$ is defined as

$$T_d \coloneqq K_d \frac{\mathrm{FW}_{50,0} I_1}{\mathrm{FW}_{50,1} I_0}, \quad (12)$$

which approaches unity in the limit of infinitesimally small opening angles.

The resulting filtered spectra for increasing opening angle are shown in *Figure 5*. These spectra are used to compute $I_1$, $\mathrm{FW}_{50}$, $\mathrm{FW}_\mathrm{K}$ and $L$, for varying $\alpha$, which are summarized in *Table 1.* To quantitatively compare the $\mathrm{FW}_{50}$ to $d\phi_{\alpha,\mathrm{Co,B}}$, a normalized energy resolution $\hat{\phi}$ is defined as $\hat{\phi} \coloneqq \mathrm{FW}_{50}/d\phi_{\alpha,\mathrm{Co,B}}$. Although $\hat{\phi}$ improves for larger opening angles, the tails of the energy distribution caused by particle interactions become more significant and eventually rise toward the peak of the spectrum near $\alpha = 2$mrad. The energy spectrum begins to significantly broaden around $\alpha = 1.5$mrad or corresponding to an input current of 11 nA, reaching $\mathrm{FW}_{50} = 19$ meV while $L$ simultaneously increases. For a user, the transmitted brightness determines the amount of current that can be focussed to a given radius in a microscope and should thus be taken into account.

Therefore, an additional dimensionless quality factor $Q$ is defined as $Q \coloneqq T_d/\hat{\phi}$, proportionally balancing the conservation of brightness with normalized energy resolution. As shown in table 1, there is an optimum for $Q$ close to 2.0 mrad. However, this $Q$ varies around its optimum, while blur due to parasytic aberrations, and thereby effective resolution loss, scale with beam angles, which is not accounted for by $Q$. Therefore, instead of simply optimizing $Q$, the design angle is biased towards a lower $\alpha = 1.5$ mrad, accounting for acceptable alignment tolerances while remaining close to the optimal $Q$ value for this design.

The simulated electron interaction effects scale with beam current, which is substantially lower for the filtered beam. Consequently, the long tails in the distribution can be suppressed using a second monochromator, which may be beneficial for some applications such as EELS. In this analysis, the effect of electron diffraction by the monochromator apertures has not yet been included, which will be separately addressed in the next section.

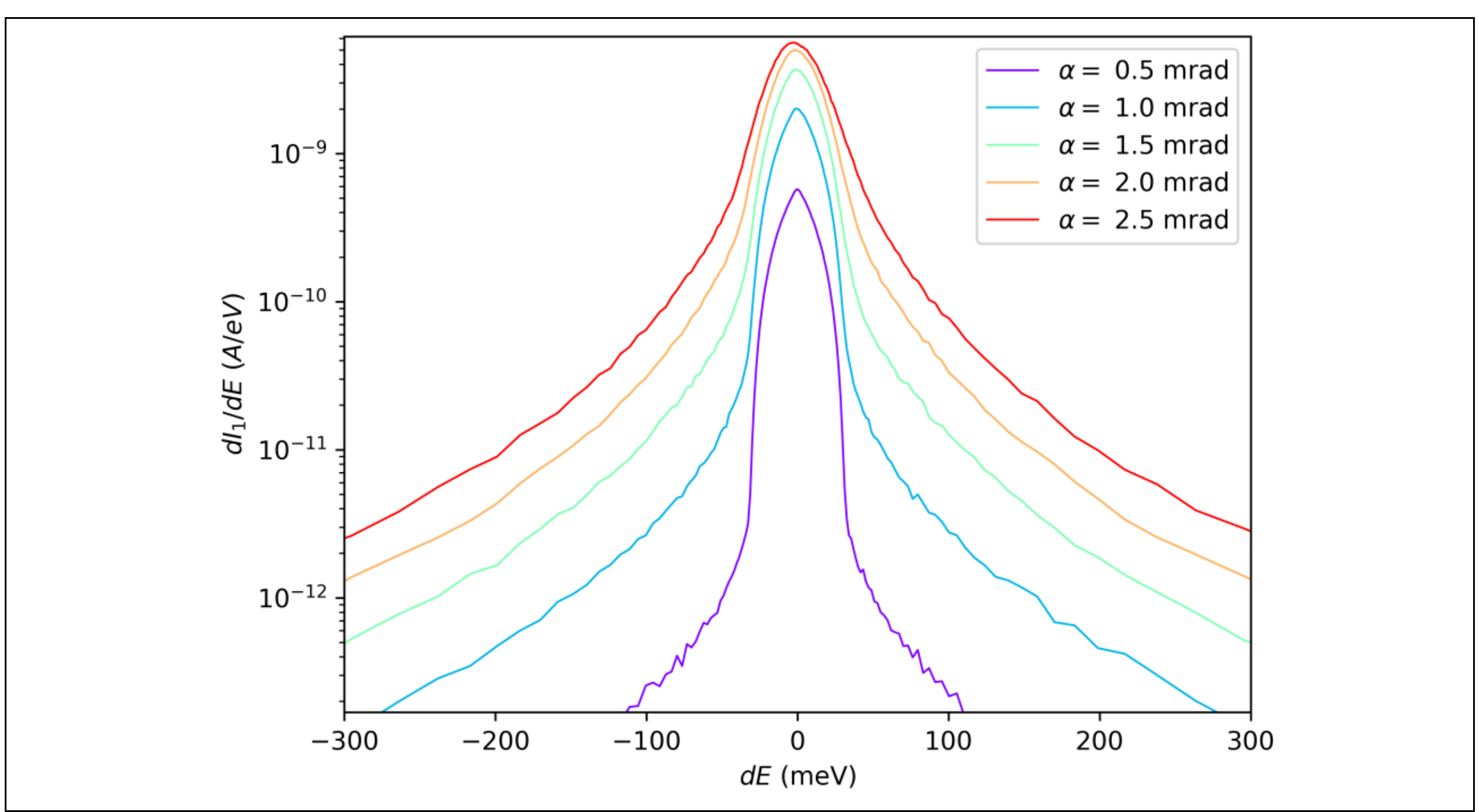

*Figure 5: (a) Filtered spectral current density $\frac{dI_1}{dE}$ shown logarithmically for varying beam opening angles $\alpha$, thereby varying the input current. The energy distribution broadens ats higher currents since the traced electrons increasingly experiencing the effects of aberrations and particle interactions.*

*Table 1: Simulated Coulomb interaction results for increasing opening angle $\alpha$. The input and output currents $I_0$ and $I_1$ broadens the spectrum, increasing both $\mathrm{FW}_{50}$ and $\mathrm{FW}_K$. At the same time, the brightness loss, $L$ increases while normalized resolution $\hat{\phi}$ increases, with an optimum for quality $Q$.*

| $\alpha$ (mrad) | $I_0$ (nA) | $I_1$ (pA) | $\mathrm{FW}_{50}$ (meV) | $\mathrm{FW}_K$ (meV) | $L$ (%) | $\hat{\phi}$ | $Q$ |
|---|---|---|---|---|---|---|---|
| 0.5 | 1.23 | 16.1 | 16 | 132 | 4 | 10.85 | 0.09 |
| 1.0 | 4.93 | 62.6 | 17 | 229 | 14 | 2.91 | 0.30 |
| 1.5 | 11.10 | 128 | 19 | 315 | 29 | 1.44 | 0.49 |
| 2.0 | 19.74 | 203 | 23 | 387 | 46 | 0.95 | 0.57 |
| 2.5 | 30.84 | 278 | 28 | 479 | 61 | 0.75 | 0.52 |

## 4.2 Diffraction and energy spectrum

The effect of diffraction on the monochromator performance is evaluated using standard wave-optical procedures [29], [30], [31]. Diffraction happens once at a farfield pupil before the monochromator, and twice at the monochromator at AP0 and AP1. This is schematically shown in *Figure 6*.

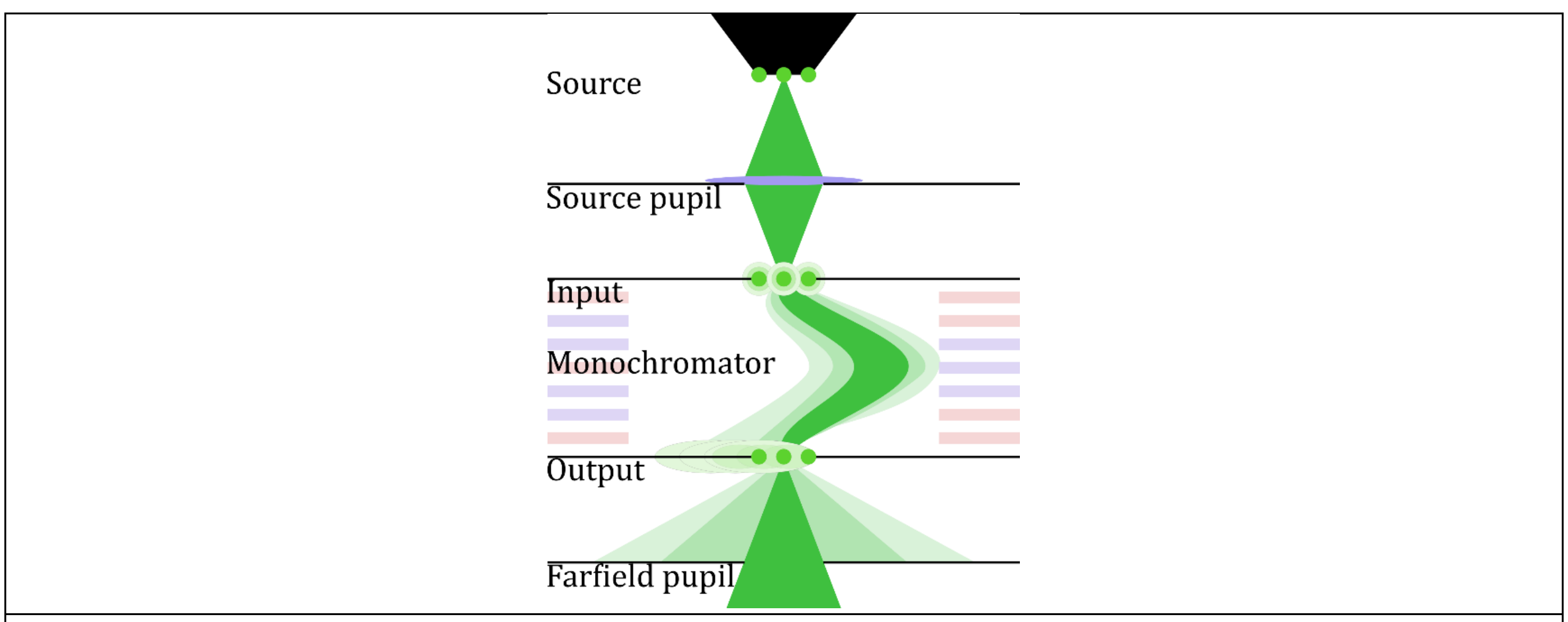

*Figure 6: Schematic layout of beam path through the monochromator and the procedure for evaluating the effects of electron diffraction on monochromator performance. The source is modeled as an ensemble of incoherent point emitters, each of which produces an Airy disk at the monochromator entrance plane. These airy disks are partially blocked by the input aperture. Then the disks are individually, coherently distorted by the aberrations of the electron-optical system before reaching the output aperture. The part of the wavefront at large beam angles is cut away by a farfield aperture. The intensities for each point emitter are individually and coherently computed and then added incoherently.*

The source is modeled as a distribution of incoherent point emitters. To estimate the amount of point emitters that need to be simulated to cover the input slit, the diameter of an airy disk that contains 50% of the current [32], $d_a = 0.54\frac{\lambda}{\alpha}$, is used where $\lambda$ is the de Broglie wavelength. The required number of simulated points is $N_p \approx \left(\frac{2r_0}{d_a}\right)^2$, arranged in golden ratio spiral [33]. The emission from each of these points is limited by a source pupil (corresponding to $\alpha = 1.5$ mrad), which is the starting plane of the diffraction simulation. Each of these points is shifted by tilting the wavefront in the pupil plane, leading to Airy disks that are clipped by the entrance slit of the monochromator. The wavefunction at the selection slit, $\psi_2$, is computed from the initial wavefunction, $\psi_1$, using Fourier propagation,

$$\psi_2(x,y) = \mathcal{F}^{-1}\{\mathcal{F}\{\psi_1(x,y)\}\exp(ik\chi)\}, \quad (13)$$

where $\mathcal{F}$ and $\mathcal{F}^{-1}$ denote the Fourier and inverse Fourier transforms, respectively, $k$ is the wavenumber, and $\chi$ is the aberration function computed from the fitted ray tracing aberration coefficients during optimization. The Fourier propagation method only assumes isoplanatic aberrations, where indices $l = m = 0$, which is a reasonable approximation after focusing. In addition, the method assumes that the beam is aberrated in the far-field, which is true when the propagation distance follows the Fraunhofer condition ($D_{FF} \gg \frac{r_0^2}{\lambda} \approx 182$ µm) at $E = 500$ eV.

The resulting point-spread functions (PSFs) for these emitters are then added incoherently to produce the spatial intensity distribution at AP1. This process is repeated for variations $dE$ around a nominal energ $E_0$, which results in the shifted transmitted intensities shown in *Figure 7*. The current transmitted through the virtual aperture, indicated by the white dashed circle, determines the transmission for a given energy $E$.

The transmitted intensities for a range of energies around $E_0$, forms a transmission spectrum $T(E)$, shown as the red curve in *Figure 8*. The simulated distributions exhibit strong asymmetry caused primarily by large coma contribution that folds high angles caused by diffraction at AP0 to distribution tails at the higher energy side of $T(E)$, suggesting inherent high assymetric broadening of the tails by diffraction. However, $T(E)$ is a projection of the distribution along $E$, while the distribution still contains a correlation $E$ and the off-axial momentum. The high energy tails of the spectrum are mostly carried by larger angles or spatial frequencies.

A farfield pupil at $\alpha = 1.5$ mrad after AP1, can suppress most of these tails, while transmitting the bulk of the central spectrum. This results in a farfield spectrum, $T_{\mathrm{FF}}$, which is also shown as the blue curve in *Figure 8*. The farfield aperture reduces the $\mathrm{FW_K}$ from 105 to 69 meV, while the $\mathrm{FW_{50}}$ remains equal to 16 meV. The maxima of $T$ and $T_{\mathrm{FF}}$, both indicated in the figure, show a minor fraction of transmission loss. This is because the probe size is not limited by diffraction at 500eV.

The effect of diffraction becomes more important when the beam energy is reduced to 5 eV, while keeping the opening angle and slit radius the same. This can be seen by the broadened intensity distribution in *Figure 7* and $T(E)$ for $E_0 = 5$ eV shown in *Figure 8*. In this case, a farfield aperture reduces the $\mathrm{FW_{50}}$ from 0.18 meV to 0.14 meV and the $\mathrm{FW_K}$ reduces from 3.54 meV to 1.01 meV. In both cases however, the spot broadening due to diffraction leads to lower transmission.

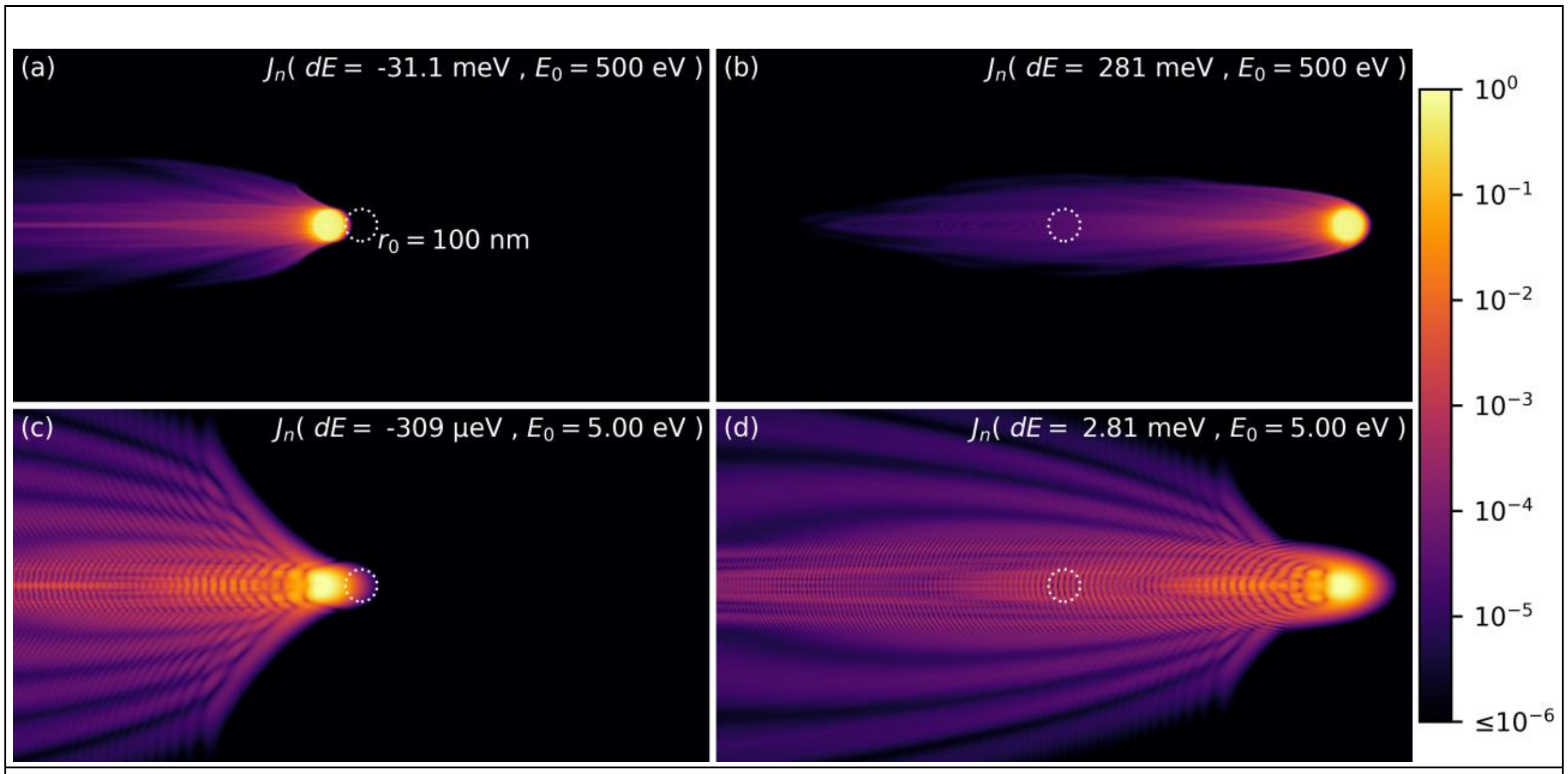

*Figure 7: Normalized logarithmic spatial current density $J_n$ as a fraction of the maximum initial density, for a wavefunction with energy offsets of (a) $dE = -31.1$ meV and (b) $dE = 281$ meV relative to a nominal beam energy of $E_0 = 500$ eV. The dashed circle indicates the perimeter of the virtual selection slit, that collects current for a given nominal energy. Figures (c) and (d) show similar results for a nominal beam energy of 5 eV.*

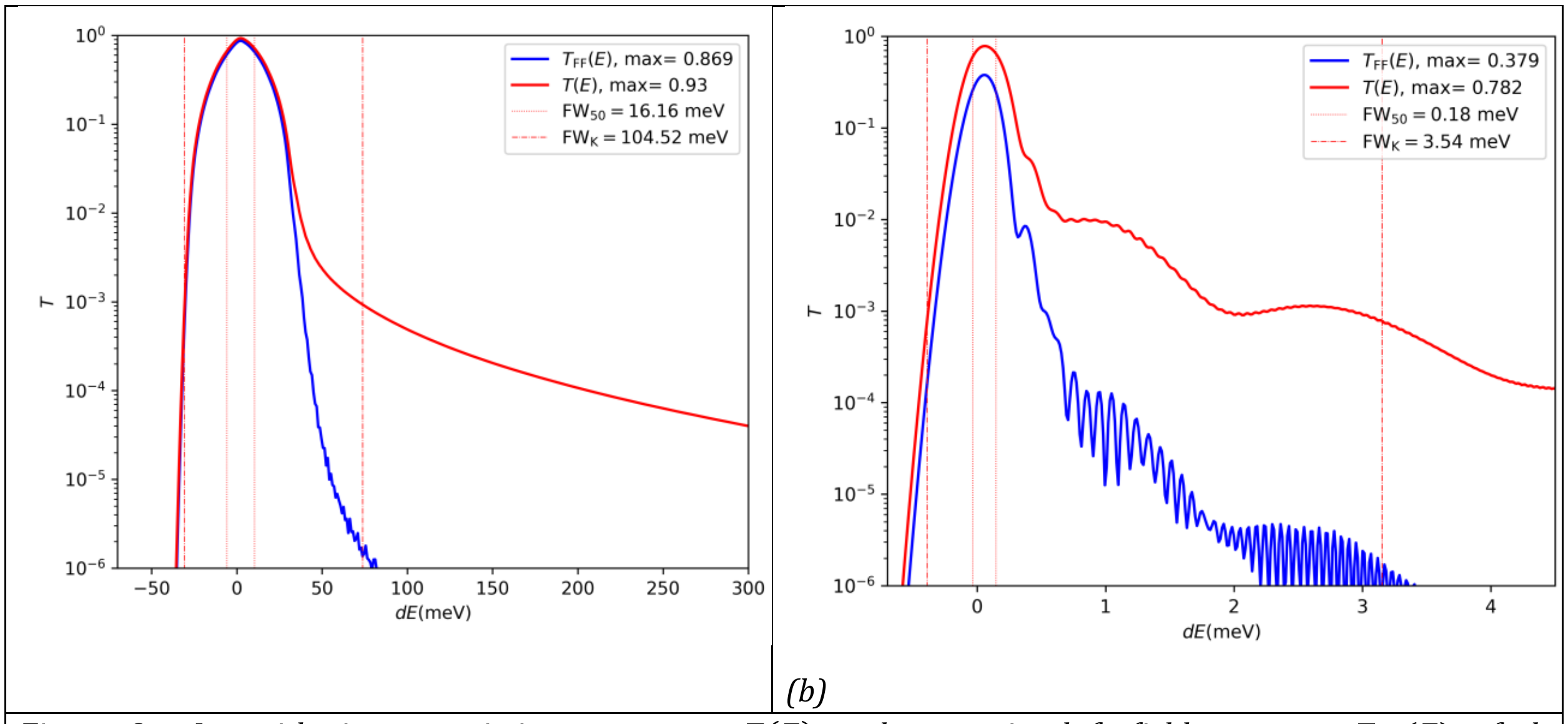

*Figure 8: Logarithmic transmission spectrum $T(E)$ and transmitted farfield spectrum $T_{FF}(E)$ of the monochromator for nominal beam energies leaving the monochromator at nominal energies (a) $E_0 = 500$ eV and (b) $E_0 = 5$ eV.*

# 5 Discussion and conclusion

We set out to design a simple electrostatic monochromator that achieves an optimal tradeoff for current and energy resolution, which is described by the theoretical approximation $d\phi_{\alpha,\mathrm{Co,B}}$. We have presented a compact design that achieves this limit. The new monochromator relies on electrodes with dimensions that can be fabricated with MEMS technology. Moreover, the design only requires 7 voltages to be supplied. We have performed an optimization of the monochromator design by first selecting a mechanical layout and then performing an iterative optimization of the electrode voltages. More design parameters and perhaps tolerancing could be incorporated in a more computationally costly optimization procedure, potentially leading to a better, higher dimensional global optimum. However, this is not expected to drastically increase the performance, since the found optimum already reaches the theoretically approximated optimal performance.

The result is a compact monochromator that achieves an energy resolution of 19 meV, comparable in magnitude to the resolution achieved in current TEMs. This is obtained with a filtered beam current of 128 pA, which is sufficient for user-friendly imaging in (LV)SEM. Moreover, the design can be tuned to diffraction limited current by lowering the nominal beam energy to achieve 0.14 meV energy resolution. The inherent assymetric tails that exist at the positive $dE$ side of the filtered energy distribution due to diffraction, which can limit the energy resolution in some EELS experiments, can be suppresed by another far-field aperture after the monochromator. The steep cutoff at the negative $dE$ side can be a major advantage in other experiments. The direction of this tail could potentially be flipped by overcorrecting the inherent coma of an electrostatic monochromator, but this would further complicate the design.

The overlapping fringe fields are expected to result in lenient tolerances for both voltage fluctuations and mechanical misalignments. Therefore, we evaluate tolerances for both voltage fluctuations and mechanical misalignments. Firstly, the voltage tolerances are based on the sensitivities $\boldsymbol{S}_i$ as introduced previously. A correlated voltage tolerance $T_{\phi,\parallel}$ and an uncorrelated voltage tolerance $T_{\phi,\perp}$ are defined as

$$T_{\phi,\parallel} = \frac{|\boldsymbol{b}|}{\sum_i |\boldsymbol{S}_i|} \qquad T_{\phi,\perp} = \frac{|\boldsymbol{b}|}{\sqrt{\sum_i |\boldsymbol{S}_i|^2}}. \tag{14}$$

The resulting electric potential drift tolerances are $T_{\phi,\parallel} = 9$ mV and $T_{\phi,\perp} = 21$ mV which are close to the theoretical single monochromator performance of 19 meV. The individual tolerances are listed in Appendix 6.2. Depending on the definition of the ground potential in the experimental implementation, the largest applied potential is the 500 V acceleration potential, which yields a relative stability of 42 ppm for the uncorrelated case and 18 ppm for the correlated case. These tolerances are of the same scale as the energy spread of the beam itself and the applied voltages are of the same magnitude as the beam energy. Therefore, a microscope that incorporates this monochromator would typically have another tolerance for the electrostatic potential that is either more strict or the same.

A similar procedure is used to evaluate mechanical alignment tolerances. In this case, small mechanical displacements replace $\partial\phi_i$ in Equation 6. The tolerances are based on imperfections that induce the aberration coefficient $C_{020001}$, commonly denoted $A_{1i}$, which corresponds to the first aberration term that cannot be compensated and is associated with the strict $10^{-5}$ rad tolerance criterium discussed in [17]. These imperfections are shifts in the $y$-direction of L0, L1 and parallel and antiparallel rotation of D1 deflector plates around the $z$-axis. The magnitude of these shifts is 5 µm and the magnitude of these rotations is 5 mrad. These misalignments introduce beam shift in the $y$-direction, which can be compensated by the integrated $y$-deflectors at D1. However, this also increases the aberration coefficient $C_{020001}$. The $x$-shifts can be corrected and are within typical alignment accuracy of several micrometers and are therefore neglected. To combine the individual tolerances to a correlated and uncorrelated single value, the rotations are converted into equivalent lateral displacements through assigning a reasonable distance of 5 mm, between the optical axis and a mechanical alignment reference for these MEMS devices.

The resulting mechanical tolerances are $T_{M,\parallel} = 53$ μm, $T_{M,\perp} = 73$ μm, which are close together because of the dominance of the L1 shift. The individual tolerances are 454 μm, 75 μm, 801 μm and 457 μm for the L0 shift, L1 shift and parallel and anti-parallel rotation of D1, respectively. It should be noted that these tolerances are only a linear approximation, as 454 μm exceeds the diameter of L0 itself, which means that nonlinear effects become significant before this limit is reached. It should be noted that the intrinsic rotation tolerance for a homogeneous deflector of $2 \times 10^{-7}$ m has been relaxed by almost three orders of magnitude, demonstrating the predicted advantange of a monochromator fully based on short, overlapping electrostatic fringe fields.

In conclusion, we have demonstrated the high performance of a new, compact, electrostatic fringe-field based electron monochromator. The monochromator achieves narrow energy resoltution with a high filtered current in combination with lenient tolerances. This promises to make high performance monochromation achievable for (LV)SEM, but also for other applications. The design presented in this manuscript can be fully realized using MEMS technology, which is currently being pursued in our lab.

# 6 Appendix

## 6.1 Dimensions

*Table A1: The mechanical dimensions of the monochromator electrodes. The electrode labels in the dimension name correspond to the labels in Figure 2.*

| Dimension name | Size (in μm) |
|---|---|
| AP0, AP1 radius | 0.1 |
| L0, L1, L2, L3 thickness | 300 |
| D0, D1, D2 thickness | 525 |
| Spacers thicknesses | 300 |
| L0, L3 radius | 200 |
| L1, L2 radius | 600 |
| D0, D2 separation distance | 600 |
| D0 positive x-offset | 50 |
| D1 separation distance | 1000 |
| D1 positive x-offset | 200 |
| D1 stigmator tip radius (and half-width) | 167 |
| D1 distance between stigmator tips | 1333 |

## 6.2 Voltages

*Table A2: The voltages applied to the monochromator electrodes. The electrode labels in the variable name correspond to the labels in Figure 2.*

| Variable name | Excitation (in V) | Further explanation | Tolerance (mV) |
|---|---|---|---|
| AP0, L0, L3, AP1 input and output acceleration | 500 | Equal to the entrance and exit acceleration potential, relative to the source. This is typically defined the reference potential here, so it is not counted as another potential supply and does not have a tolerance. | - |
| D0, D2 average | 337.888 | | 167 |

| D0, D2 deflection | 121.927 | +Deflection/2 for the deflector positioned in the positive x direction | 42 |
|---|---|---|---|
| L1, L2 acceleration | 77.247 | Acceleration potential at the surface of L1 and L2 | 46 |
| D1 average | 136.066 | | 63 |
| D1 deflection | -167.014 | +Deflection/2 for the deflector positioned in the positive x direction | 35 |
| D1 stigmation | -7.505 | -stigmation/2 for the deflector plates, +stigmation/2 for the tips. | 90 |
| D1 y-deflection | 0. | Can be used to align the beam in the y-direction. | 182 |

## 6.3 Aberrations

Table A3: Aberration values after optimization.
*This is supposed to be 0 due to the symmetry of the design. The finite scale it does have is a measure of the computational inaccuracies of the meshing, solving charge densities, tracing and aberration fitting. With a beam angle of 1.5 mrad, the offsets produced by A1i are < 1 nm. These imaginary aberrations are not used for optimization costs, since there is no degree of freedom available in optimization to correct for it. dy is insignificant compared to any feasible alignment accuracy for this array of macroscopic shapes, and should thus simply be corrected for.
**Though the polynomial corresponds to an ordinary drift space, the polynomial is $2(x' + iy')$, so there is a factor 2 between this polynomial and intuitive defocus.
***The negative sign is due to the convention of image offsets as a function of object angles. Division by M results in a positive valued, conventional object side spherical aberration.

| Abbreviation $(j, k, l, m, c, r)$ | Name | Symbol | Value (in $\mathrm{m}^{1-l-m}$) |
|---|---|---|---|
| 0,1,0,0,0,0 | x-displacement | dx or B0r | -2.4591722367234867e-09 |
| 0,1,0,0,0,1 | y-displacement* | dy or B0i | -7.697005194683075e-08 |
| 0,1,1,0,0,0 | Magnification | M | -1.0004413556000995 |
| 0,1,0,1,0,0 | Asymmetric magnification | C010100 | -0.00030865808253706996 |
| 0,2,0,0,0,0 | Twofold-astigmatism | A1r | -9.030347586503942e-07 |
| 0,2,0,0,0,1 | Diagonal twofold-astigmatism* | A1i | -1.9032712936483903e-07 |
| 1,1,0,0,0,0 | Defocus** | C1 | +1.186445130238927e-06 |
| 0,3,0,0,0,0 | Threefold-astigmatism | A2 | -0.0022289225205611956 |
| 1,2,0,0,0,0 | Axial coma (or spherical displacement) | B2 | -0.008995796211786202 |
| 0,4,0,0,0,0 | Fourfold-astigmatism | A3 | +0.02177185165352178 |
| 1,3,0,0,0,0 | Second order twofold-astigmatism | S3 | +0.03899683751180235 |
| 2,2,0,0,0,0 | Spherical aberration*** | C3 | -0.01305718717559376 |
| 0,1,0,0,1,0 | Chromatic x-dispersion | dxc1 | +0.00319758263139424 |
| 1,1,0,0,1,0 | Chromatic defocus | Cc | 0.012293059810348896 |